%Paper: hep-th/9309073
%From: t14269@mail.ncku.edu.tw
%Date: Mon, 13 Sep 93 21:38:18 CST
%Date (revised): Wed, 22 Jun 94 09:09:49 CST

\input phyzzx
\font\bigmath=cmex10
\font\medium=cmbx10 scaled \magstep1
\font\large=cmbx10 scaled \magstep2
\vsize 23.5 cm
\hsize 16.5 cm
\hoffset 0.1 cm
\voffset 0 cm
\line{\hfill\hbox{NCKU-HEP/93-09}}
\line{\hfill\hbox{Revised Nov. 30, 1993}}
\line{\hfill\hbox{Title changed}}
\vskip 3 cm
\centerline{\large A Non-Principal Value Prescription for the Temporal Gauge}
\vskip 3cm
\centerline{Kuo-Cheng Lee}
\vskip 0.2cm
\centerline{\it and}
\vskip 0.2cm
\centerline{Su-Long Nyeo}
\vskip 0.4 cm
\address{Department of Physics, National Cheng Kung University}
\address{Tainan, Taiwan 701, Republic of China}
%\vfill \eject

\noindent
Running Title: Temporal Gauge

\abstract
\noindent
A non-principal value prescription is used to define the spurious
singularities of Yang-Mills theory in the temporal gauge.
Typical one-loop dimensionally-regularized temporal-gauge integrals
in the prescription are explicitly calculated, and a regularization for
the spurious gauge divergences is introduced.
The divergent part of the one-loop self-energy is shown to be local
and has the same form as that in the spatial axial gauge with the
principal-value prescription.
The renormalization of the theory is also briefly mentioned.
\filbreak
\vfill \eject

\noindent{\bf\medium 1. Introduction}

\noindent
Quantization of constrained systems such as gauge theories is a very delicate
problem. In particle physics, one has been interested in quantizing gauge
theories in noncovariant gauges [1,2] like the general axial gauges, which
can be classified by constant four-vectors.  In these gauges, the theories
enjoy the possibility of the decoupling of the Faddeev-Popov ghosts.  However,
these theories possess residual symmetries, which lead to difficulties in
defining the propagators for the gauge fields.  In order to specify the
propagators, one has to define boundary conditions for the gauge fields in
the spatial-axial gauge, but to define initial conditions for the gauge
fields in the temporal gauge [2].

It is generally known that among the noncovariant gauges, the temporal gauge
is the most complicated and cumbersome gauge choice, even from the point of
view of the perturbation expansion.  It remains a difficult problem to
define useful Feynman rules for perturbative calculations in the temporal
gauge.  Because Feynman rules are Green functions, they are distributions and
integrations with respect to loop variables may not commute [3].
For example, integrations in gauge theories
in the spatial axial gauge with the principal-value prescription
do not necessarily
commute.  Moreover, for applications in particle physics, it is useful to use a
renormalizable gauge formalism for calculating radiative corrections.  Thus,
one should investigate the renormalization structure of any viable gauge
formalism, which may find applications in particle physics.  Besides, with the
importance of the nontrivial topological vacuum structure of Yang-Mills theory,
which has been investigated [4] in the temporal gauge, a perturbative study
can be useful to nonperturbative effects in processes involving hadrons.
%although perturbative calculations continue to play a major role.

In this paper, we shall study the renormalization structure of Yang-Mills
theory quantized in the temporal gauge.  In particular, we shall treat the
unphysical singularities in the polarization sum in the gauge propagator by a
non-principal value prescription, which enjoys the property of being
commutative with
respect to integration.  Such a prescription has been introduced many years ago
[5,6] but has not been analyzed or applied to particle physics. We organize
this paper as follows.  In section 2, we shall introduce the non-principal
value prescription and use it, in section 3, to calculate two typical
dimensionally-regularized temporal-gauge integrals for the one-loop gluon
self-energy.  In section 4, we shall briefly mention the renormalization of
the gauge theory in this formalism.  Finally, we shall provide a short section
for discussion.
\vskip 1cm

\noindent {\bf\medium 2. The Non-Principal Value Prescription}

\noindent
The temporal gauge is defined by the condition
$n_\mu A_{\mu}^a = A_{0}^a = 0$, where $A_{\mu}^a$ is the gauge potential, and
$n_\mu = (1,0,0,0)$ a constant temporal four-vector. In this gauge, the
propagator has a spurious double pole at $q_{_{0}} = 0$ and reads $(i,j =
1,2,3)$
$$
G_{ij}^{ab}(q) = {{i\delta^{ab}}\over{q^2 + i\epsilon}}\left\{
\delta_{ij} + {{q_iq_j}\over{q_{_{0}}^2}}\right\},\,\,
G_{0i}^{ab}(q) = G_{i0}^{ab}(q) = G_{00}^{ab}(q) = 0,\eqno(1)
$$
or, in covariant form,
$$
G_{\mu\nu}^{ab}(q) = {{i\delta^{ab}}\over{q^2 + i\epsilon}}\left\{
- \delta_{\mu\nu} + {{(q_\mu n_\nu + q_\nu n_\mu)}
\over{q\cdot n}}
- {{q_\mu q_\nu}
\over{(q\cdot n)^2}}\right\}\,\,,\,\epsilon > 0\,,\eqno(2)
$$
where we use a $(+,-,-,-)$ metric.  We note that propagator (2)
has a simple pole and a double pole at $q_{_{0}} = 0$.  In calculations,
these unphysical poles should be circumvented with prescriptions in a way that
physical quantities should not depend on the
prescriptions and on which of the above propagators one uses.

To date, many prescriptions with regularizations for the spurious poles
have been introduced.  We shall consider a first-order type of prescription,
which is defined for the simple pole, with its square for the double pole.
For example, we may consider a prescription given by the replacement [7]:
$$
\eqalignno{
{1\over q_{_{0}}} &\rightarrow {1\over {q_{_{0}}-\alpha|\vec q|}}
= {
{q_{_{0}} + \alpha|{\vec q}|}\over
{{q_{_{0}}^2} - \alpha^2{\vec q}^2}}\cr
&{}\cr
&\rightarrow \lim_{\epsilon \rightarrow 0}
\left[
{{q_{_{0}} + \alpha|\vec q|}\over{{q_{_{0}}^2} - \alpha^2{\vec q}^2 +
i\epsilon}} \right]\,,\epsilon > 0\,,
&(3)\cr}
$$
and the double pole is given by
$$
\eqalignno{
{1\over{{q_{_{0}}}}^{\!\!2}}
&\rightarrow \lim_{\epsilon \rightarrow 0}
{\left[
{{q_{_{0}} + \alpha|\vec q|}\over{{q_{_{0}}^2} - \alpha^2{\vec q}^2 +
i\epsilon}} \right]}^2,\,\epsilon > 0\,,
&(4)\cr}
$$
where $\epsilon$ specifies how the $q_{_0}$ integration is to be carried out,
and the limit $\epsilon\rightarrow 0$ is taken only after the integration.  The
parameter $\alpha$, whose limit $\alpha\rightarrow 0$ should be taken only
after a calculation of a given loop order in perturbation theory,
depends on the sign of $q_{_{0}}$ and is used for
regularizing the divergences arising from the gauge singularities.  Note
that the prescription for the double pole is non-negative.

In this paper, we shall consider a non-principal value prescription defined by
replacing the simple pole by
$$
\eqalignno{
{1\over q_{_{0}}} &\rightarrow {1\over{[q_{_{0}}]}}  \equiv
\lim_{\epsilon \rightarrow 0} \left[
{{q_{_{0}} }\over{{q_{_{0}}^2} + i\epsilon}} \right]\,,\epsilon > 0\,.
&(5)\cr}
$$
Then the double pole is given by
$$
{1\over{q_{_{0}}}^{\!\!2}} \rightarrow {1\over{[q_{_{0}}^2]}}
\equiv {1\over{[q_{_{0}}]^2}} = \lim_{\epsilon \rightarrow 0}
{\left[
{{q_{_{0}} }\over{{q_{_{0}}^2} + i\epsilon}} \right]}^2\,,\epsilon > 0\,.
\eqno(6)
$$
Definitions (5) and (6) can be seen as the reduced forms of (3) and (4) with
$\alpha = 0$.
The two poles in (6), $q_0^\pm = \mp \sqrt{i\epsilon}$,
merge with one another as $i\epsilon\rightarrow 0$.  These
pinching poles are the problem of using the temporal gauge.
There is no problem with (5), which has essentially a simple pole.
We emphasize that the ``$i\epsilon$" in (5) and (6) only provides us with
a definition for $q_0$ integration.  Dimensionally-regularized integrals
with the spurious poles of (6) are not regularized.
We should expect that dimensional regularization does not regularize
the gauge singularities, since gauge singularities and space-time
dimensionality are unrelated, with space-time dimensionality being intimately
related to the physical structure of a theory.
The need for a regularization for integrals with the double gauge pole
may be seen in the following section.

The above definitions, (5) and (6), are also seen to be related
to those introduced by Leibbrandt [8], who considers the temporal
gauge $n_\mu A_{\mu}^a = 0$ with a constant temporal vector
$n_\mu \equiv (n_0,{\bf n}_\bot, n_3 = -i|{\bf n}_\bot|)$.
For his definitions, a dual temporal vector $n_\mu^{\rm dual}
\equiv (n_0,{\bf n}_\bot,i|{\bf n}_\bot|)$ is also introduced.
The basic one-loop temporal gauge integrals are local,
but the one-loop gluon self-energy is nonlocal in the external momentum.
For real ${\bf n}_\bot$ and $|{\bf n}_\bot| \rightarrow 0$,
his definitions for the simple and double poles
reduce algebraically to (5) and (6) above. In [8],
the two constant temporal vectors
play the role of a regularization for the gauge divergences.
%and no regularization is needed.  However, we see that some of the
%basic one-loop integrals derived with the definitions in
%[8] do not reduce to the corresponding integrals derived with (5) and
%(6) in the pure temporal limit ${\bf n}_\bot \rightarrow 0$.
%This is not too surprising, since
%different regularizations can lead to different gauge-dependent
%results.

It is perhaps tempting to define the double pole as
$$
{1\over{q_{_{0}}}^{\!\!2}}
\equiv \lim_{\epsilon \rightarrow 0}
{\left[
{1\over{{q_{_{0}}^2} + i\epsilon}} \right]}\,,\epsilon > 0\,.
\eqno(7)
$$
This definition was introduced long time ago [5,6] and can be the same as (6)
if a particular regularization is used.  In general, definitions (6) and (7)
do not necessarily lead to finite temporal-gauge integrals but merely
provide calculational procedures.  The divergences arising from the gauge
singularities,
as we shall see, require also a regularization.  In short, a prescription
amounts to providing definitions for the spurious gauge
poles and may or may not include a regularization.

In the following section, we shall employ the above definitions for the poles
in the one-loop integrals.  We look for a regularization scheme such that
the definitions enjoy certain useful properties.  They are (under integral
signs):
$$
{q_{_{0}}\over{[q_{_{0}}]}} = 1\,, \eqno(8)
$$
$$
{q_{_{0}}\over{[q_{_{0}}^2]}} = {1\over{[q_{_{0}}]}}\,,
\eqno(9)
$$
$$
{{q_{_{0}}^2}\over{[q_{_{0}}^2]}} =  1\,.
\eqno(10)
$$
In addition, the algebraic identity
$$
{1\over{q\cdot n(p - q)\cdot n}}
= {1\over{p\cdot n}}\left[
{1\over{q\cdot n}} +
{1\over{(p - q)\cdot n}}\right]\,,\,\, p\cdot n \ne 0\,\,\,,
\eqno(11)
$$
if also satisfied, would be useful for simplifying calculations.
\vskip 1cm

\noindent{\bf\medium 3. Sample Calculations}

\noindent
The purpose of this section is to outline the calculational procedure for the
one-loop dimensionally-regularized temporal-gauge integrals and to observe
the gauge divergence problem of temporal-gauge theories.  At the one-loop
level, we shall need a regularization for regularizing the gauge divergences of
the integrals.  For the usual ultra-violet divergences that we are interested
in, we employ dimensional
regularization with complex space-time dimensionality $2\omega$.

First consider the following integral with the spurious simple pole:
$$
I = \int {{d^{2\omega}q}\over{(p - q)^2q\cdot n}}\,.
\eqno(12)
$$
Using prescription (5) for the simple pole and Feynman's parametrization
formula, we get
$$
\eqalignno{
I &= \int_0^1 d\alpha\int {{d^{2\omega}q\,\,\,\,\,\,
q_0}\over{\left[\alpha\left((p - q)^2 + i\epsilon
\right) + (1 - \alpha)\left(q_0^2 +
i\epsilon\right)\right]^2}}\cr
&= \int_0^1 d\alpha\int d^{2\omega - 1}{\vec q}\,\int_{-\infty}^{+\infty}
{{dq_0\,\,\,
(q_0 + \alpha p_0)}\over{\left[q_0^2 +
\alpha\left(1 - \alpha\right)p_0^2 - \alpha\left({\vec p} - {\vec q}\right)^2
+ i\epsilon\right]^2}}\,.&(13)\cr}
$$
The $q_0$ integral is carried out first and leads to
$$
I = {{i\pi p_0}\over2}\int_0^1 {{d\alpha}\over{\sqrt{\alpha}}}\int
d^{2\omega - 1}{\vec q}\,\,
\left[p_0^2(\alpha - 1)
+ ({\vec p} - {\vec q})^2\right]^{-3/2}\,.\eqno(14)
$$
To perform the ${\vec q}$ integral, we use the integral representation
$$
{1\over{A^n}} = {1\over{\Gamma(n)}}\int_0^{\infty}dx\, x^{n-1}\,
e^{-xA}\,, A > 0\,,\eqno(15)
$$
and get, taking $p_0^2 < 0$,
$$
I = ip_0\int_0^1 d\alpha\int_0^\infty dx\sqrt{{\pi x}\over{\alpha}}
\int d^{2\omega - 1}{\vec q}\,
e^{-x\left[p_0^2(\alpha - 1) + ({\vec p} - {\vec q})^2\right]}\,.\eqno(16)
$$
Finally, we obtain
$$
\eqalignno{
I &= i\pi^\omega\int_0^1 {{d\alpha}\over{\sqrt{\alpha}}}
\int_0^\infty dx\,x^{1 - \omega}\, e^{-xp_0^2(\alpha - 1)}\cr
&= i\pi^\omega \Gamma(2 - \omega)p_0^{2\omega - 3}\int_0^1 d\alpha
\alpha^{-1/2}(\alpha - 1)^{\omega - 2}\cr
&= {{2p\cdot n}\over{n^2}}{\overline I} \,\,,\,\,
{\overline I} \equiv i\pi^2\Gamma(2 - \omega)\,\,,\,
\omega \rightarrow 2\,.&(17)\cr}
$$
This result has the same form as that for the corresponding spatial-axial gauge
integral calculated with the principal-value prescription.

Next we turn to an integral with the double pole:
$$
J = \int {{d^{2\omega}q}\over{(p - q)^2(q\cdot n})^2}\,.
\eqno(18)
$$
Using definition (6) for the spurious double pole and Feynman's
parametrization formula, we get
$$
\eqalignno{
J &= 2\int_0^1 d\alpha(1 - \alpha)\int d^{2\omega - 1}{\vec q}\,
\int_{-\infty}^{+\infty} {{dq_0\,\,\,
(q_0 + \alpha p_0)^2}\over{\left[q_0^2 +
\alpha\left(1 - \alpha\right)p_0^2 - \alpha\left({\vec p} - {\vec q}\right)^2
+ i\epsilon\right]^3}}\cr
&= {{\pi i}\over2}\int_0^1 {{d\alpha}\over{\sqrt{\alpha}}}(1 - \alpha)
\int d^{2\omega - 1}{\vec q}\,
\left[-{{3p_0^2}\over{2C^5}} + {1\over{2\alpha C^3}}\right]\,,&(19)\cr
}
$$
where $C^2 = p_0^2(\alpha - 1) + ({\vec p} - {\vec q})^2$.
The ${\vec q}$ integral is performed by using (15).  The first term in square
bracket in (19) leads to a finite integral and is omitted here.
For $\omega \rightarrow 2$, integral (19) reduces to
$$
\eqalignno{
J &= -i\pi^\omega\Gamma(2 - \omega)\left(p_0^2\right)^{\omega - 2}{1\over2}
\int_0^1 d\alpha \alpha^{-3/2}(\alpha - 1)^{\omega - 1}\cr
&\approx -{\overline I} +
{{\overline I}\over2}\int_0^1d\alpha\alpha^{-3/2}\,,\,\,
\omega \rightarrow 2\,.
&(20)\cr
}
$$
The integral in the Feynman parameter diverges, independent of $\omega$.
Therefore, we need a regularization, and the simplest regularization is to
set
$$
\int_0^1d\alpha\alpha^{-3/2} = -2\,.\eqno(21)
$$
This result is obtained by considering the following complex-valued function:
$$
f(z) = \int_0^1d\alpha\alpha^{z}\,,\eqno(22)
$$
which diverges for real z $\le -1$.  We may analytically
continue it to real z $ < -1$:
$$
f(z) \,\,\longrightarrow\,\,{\widetilde f}(z) = {1\over{z + 1}}\,.\eqno(23)
$$
It has a pole at $z = -1$ and ${\widetilde f}(-3/2) = -2$.  Hence we obtain
$$
J = {-2\over{n^2}}{\overline I}\,,\eqno(24)
$$
which has the same form as that for the corresponding spatial axial-type
integral
calculated with the principal-value prescription.  Other temporal-gauge
integrals required for the one-loop gluon self-energy are calculated and listed
in the appendix.  We see that the temporal-gauge integrals requiring a
regularization like (21) have the spurious double pole.
Integrals having the spurious simple pole do not exhibit
gauge divergences and need no regularization.  We observe that the use of (7)
leads to the same results as those from using (6), provided that we define the
only divergent integral by (21).  The identities (8), (9), (10) and the
algebraic decomposition formula (11) are satisfied if the same
regularization scheme defined by (21) is used.

\vskip 1cm

\noindent{\bf\medium 4. Renormalization}

\noindent
In this section, we shall investigate the renormalization of the temporal
gauge theory in the non-principal value formalism.
We shall need to calculate the one-loop
gluon self-energy.  Since we are interested in simple Feynman rules with the
decoupling of the Faddeev-Popov ghosts, we shall work with the equivalence
theorem [5,9].

This theorem is based on the required property that the gauge propagator be
time-translation invariant.
Let the time-translation invariant gauge propagator be:
$$
G_{\mu\nu}^{ab}(q) = {{i\delta^{ab}}\over{q^2 + i\epsilon}}\left(
-\delta_{\mu\nu} + a_\mu(q)q_\nu- a_\nu(-q)q_\mu\right)\,,\eqno(25)
$$
where $a_\mu(q)$ is an arbitrary function related to the gauge choice.
Let the ghost-gluon-ghost vertex be:
$$
{{\Gamma_\mu}^{\!\!\!abc}}(q) =
- gf^{abc}\left(\left[(a\cdot q) - 1\right]q_\mu - q^2
a_\mu(q)\right)\,,\eqno(26)
$$
and let the ghost propagator be:
$$
G(q) = {{-i}\over{{q^2 + i\epsilon}}}\,,\,\,\epsilon > 0\,,\eqno(27)
$$
with $g$ being the coupling constant, $q_\mu$ the outgoing ghost's
momentum.  Then the S-matrix is independent of the gauge-dependent
function $a_\mu(q)$.  We should mention that definitions
(26) and (27) are related, since they always appear in pairs in diagrams
with ghost loops.

{}From this theorem, we get for the temporal gauge
$$
a_\mu(q) = {{n_\mu}\over{q\cdot n}} - {{q_\mu}\over{2(q\cdot
n)^2}}\,,\eqno(28)
$$
$$
{{\Gamma_\mu}^{\!\!\!abc}}(q) =
gf^{abc}{{q^2n_\mu}\over{q\cdot n}}\,.\eqno(29)
$$
We observe that the ghost-gluon-ghost vertex (29) has a $q^2$ factor, which
cancels the $q^2$ dependence of the ghost propagator (27)
in diagrams with ghost loops.
Thus, the pair (27) and (29) is seen to be identical to
the pair of Feynman rules for the ghost propagator and the ghost-gluon-ghost
vertex
obtained by the Faddeev-Popov gauge-fixing procedure. Using the procedure,
we have a ghost-gluon-ghost vertex that is proportional to $n_\mu$ and a
ghost propagator that is inversely proportional to $q\!\cdot\!n$.
It is easy to show that in dimensional regularization
the one-loop ghost diagram vanishes in the non-principal value prescription.
Therefore, the calculation of the one-loop gluon self-energy requires
considering one diagram with an internal gluon loop.

The calculation of the divergent part of the one-loop gluon self-energy is
straightforward but tedious and yields
$$
i{\Pi_{\mu\nu}}^{\!\!\!\!\!\!ab}(p) =
{{11g^2C_A}\over{3(2\pi)^{2\omega}}}\delta^{ab}(p^2\delta_{\mu\nu}
- p_\mu p_\nu){\overline I}\,,\eqno(30)
$$
where $C_A = N$ for $SU(N)$ gauge group.  We observe that the self-energy
is transverse and independent of the temporal vector $n_\mu$.  Thus, the
renormalization structure is very simple in this formalism and has the same
form as that in the spatial axial gauge calculated with the principal-value
prescription.  The finite part of the self-energy
in the temporal gauge in the non-principal value formalism could
depend on the temporal vector $n_\mu$,
since gauge self-energy is a gauge dependent quantity.

Finally, we note that many authors have considered time-translation
noninvariant propagators [10].  The renormalizations of the gauge
theories with these propagators have yet to be investigated.

\vskip 1cm

\noindent{\bf\medium 5. Discussion}

\noindent
Among the noncovariant gauges, the temporal gauge is generally known to be
the most complicated and cumbersome gauge choice.  In this work, we have
considered a non-principal value prescription as a choice for defining
the spurious
temporal-gauge poles.  This prescription provides us with a calculational
procedure for the dimensionally-regularized temporal-gauge integrals.  In
calculating the integrals, we first transform the divergences
arising from
the gauge singularities into divergences in the Feynman integration
parameters.  These divergences are seen not to be regularized by dimensional
regularization and hence a regularization is needed.
By using the regularization scheme (21), we have showed that the results for
the integrals have the same forms
as those for the corresponding spatial-axial gauge integrals with the
principal-value prescription.  Within the regularization scheme, the basic
integrals are seen to obey the power-counting rule.
We see that the identity (11) can be used
to simplify integrals.  The divergent part of one-loop gluon self-energy
is local in the external momentum and has the same simple structure
as that in the spatial-axial gauge in the
principal-value formalism.  The renormalization structure of the temporal-gauge
theory in the non-principal value prescription is seen to be very simple.

Finally, we should mention
that our definitions, (5) and (6), are seen to be related
to those introduced by Leibbrandt [8].
In Leibbrandt's definitions for the simple and double poles,
the constant temporal vector and its dual temporal vector
play the role of a regularization,
and no regularization is needed for the gauge divergences.
We observe that certain
one-loop integrals derived with the definitions in
[8] do not reduce to the corresponding integrals derived with (5) and
(6) in the limit of vanishing spatial components of the constant
temporal vectors.  The one-loop gluon self-energy calculated in [8]
is transverse but nonlocal in the external momentum.
\vskip 1cm

\centerline{\bf\medium Acknowledgment}

\noindent
This work is supported by the National Science Council of the Republic
of China under contract number NSC 82-0208-M006-71.

\endpage
\centerline{\bf Appendix}

Here we list the ultra-violet divergent parts of the dimensionally-regularized
temporal-gauge integrals in the non-principal value prescription with
divergences
denoted by ${\overline I} \equiv i\pi^2\Gamma(2 - \omega)$.  These integrals,
which are the same as the corresponding spatial-axial gauge integrals ($n^2
= - 1$) with the principal-value prescription, are used for the evaluation of
the one-loop gluon self-energy.
\vskip 0.3cm
$$
\eqalignno{
&\int {{d^{2\omega}q}\over{(p - q)^2q\cdot n}}
= {{2p\cdot n}\over{n^2}}{\overline I}{\hskip 11.5cm}
&(A.1)\cr
&{}\cr
&{}\cr
&\int {{d^{2\omega}q\,q_\mu}\over{(p - q)^2q\cdot n}} =
{2\over{n^2}}\left[{p\cdot n}p_\mu
- {{(p\cdot n)^2n_\mu}\over{n^2}}\right]{\overline I}
&(A.2)\cr
&{}\cr
&{}\cr
&\int {{d^{2\omega}q}\over{q^2(p - q)^2q\cdot n}} = {\rm finite}
&(A.3)\cr
&{}\cr
&{}\cr
&\int {{d^{2\omega}q\, q_\mu}\over{q^2(p - q)^2q\cdot n}} =
{{n_\mu}\over{n^2}}{\overline I}
&(A.4)\cr
&{}\cr
&{}\cr
&\int {{d^{2\omega}q\, q_\mu q_\nu}\over{q^2(p - q)^2q\cdot n}} =
{1\over{2n^2}}\left[p\cdot n\delta_{\mu\nu} + p_\mu n_\nu + p_\nu n_\mu
- {{2p\cdot nn_\mu n_\nu}\over{n^2}}\right]{\overline I}
&(A.5)\cr
&{}\cr
&{}\cr
&\int {{d^{2\omega}q}\over{(p - q)^2(q\cdot n)^2}} =
{{-2}\over{n^2}}{\overline I}
&(A.6)\cr
&{}\cr
&{}\cr
&\int {{d^{2\omega}qq_\mu}\over{(p - q)^2(q\cdot n)^2}} =
{2\over{n^2}}\left[{{2p\cdot nn_\mu}\over{n^2}} - p_\mu\right]
{\overline I}
&(A.7)\cr
}
$$
\vskip 0.2cm
$$
\eqalignno{
\int {{d^{2\omega}q\,q_\mu q_\nu}\over{(p - q)^2(q\cdot n)^2}} =
&{{2}\over{n^2}}\left[- p_\mu p_\nu
+ {{2p\cdot n}\over{n^2}}\left(p_\mu n_\nu
+ p_\nu n_\mu\right)\right.{\hskip 6cm}\cr
&+ \left.{{(p\cdot n)^2}\over{n^2}}\delta_{\mu\nu}
- {{4(p\cdot n)^2}\over{n^4}}n_\mu n_\nu\right]{\overline I}
&(A.8)\cr}
$$
$$
\eqalignno{
&\int {{d^{2\omega}q}\over{q^2(p - q)^2(q\cdot n)^2}} =
{\rm finite}{\hskip 10.8cm}
&(A.9)\cr
&{}\cr
&{}\cr
&\int {{d^{2\omega}q\,q_\mu}\over{q^2(p - q)^2(q\cdot n)^2}} = {\rm finite}
&(A.10)\cr
&{}\cr
&{}\cr
&\int {{d^{2\omega}q\,q_\mu q_\nu}\over{q^2(p - q)^2(q\cdot n)^2}} =
{1\over{n^2}}\left[{{2n_\mu n_\nu}\over{n^2}} - \delta_{\mu\nu}\right]
{\overline I}
&(A.11)\cr
}
$$
$$
\eqalignno{
\int {{d^{2\omega}q\,q_\mu q_\nu q_\sigma}\over{q^2(p - q)^2(q\cdot n)^2}}
= &-{1\over{2n^2}}
{\raise17pt\hbox{\bigmath\char'042}}
p_\mu\delta_{\nu\rho}
+ p_\nu\delta_{\mu\rho} + p_\rho\delta_{\mu\nu}{\hskip 6.5cm}\cr
&- 2p\cdot n
\left(n_\mu\delta_{\nu\rho} + n_\nu\delta_{\mu\rho}
+ n_\rho\delta_{\mu\nu}\right)\cr
&- 2\left(p_\mu n_\nu n_\rho + p_\nu n_\mu n_\rho
+ p_\rho n_\mu n_\nu\right)\cr
&+ {{8p\cdot n}\over{n^2}}n_\mu n_\nu n_\rho
{\raise17pt\hbox{\bigmath\char'043}}
{\overline I}
&(A.12)\cr}
$$
\endpage

\frenchspacing

\ref{G. Leibbrandt, Rev. Mod. Phys. {\bf 59} (1987) 1067.}
\ref{A. Bassetto, G. Nardelli and R. Soldati,
     {\it Yang-Mills Theories in Algebraic Non-Covariant Gauges}, World
     Scientific, Singapore, 1991.}
\ref{T.T. Wu, Phys. Lett. {\bf 71B} (1977) 142; Phys. Rep. {\bf 49} (1979)
     245.}
\ref{M. Creutz and T.N. Tudron, Phys. Rev. {\bf D16} (1977) 2978;
     M. Creutz, I.J. Muzinich and T.N. Tudron, {\it ibid.} {\bf D19} (1979)
     531; F.M. Saradzhev and V.Ya. Faynberg, in {\it Solitons and Instantons,
     Operator Quantization}, V.L. Ginzburg (ed.), Nova Science, Commack,
     1987.}
\ref{R. Mills, Phys. Rev. {\bf D3} (1971) 2969.}
\ref{K. Huang, {\it Quarks, Leptons \& Gauge Fields}, World Scientific,
     Singapore, 1982.}
\ref{S.-L. Nyeo, Z. Phys. {\bf C52} (1991) 685.}
\ref{G. Leibbrandt, Nucl. Phys. {\bf B310} (1988) 405.}
\ref{H. Cheng and E.-C. Tsai, Phys. Rev. Lett. {\bf 57} (1986) 511.}
\ref{S. Caracciolo, G. Curci and P. Menotti, Phys. Lett. {\bf 113B} (1982) 311;
H.D. Dahmen, B. Scholz and F. Steiner, Phys. Lett. {\bf 117B} (1982) 339;
S.C. Lim, Phys. Lett. {\bf 149B} (1984) 201;
ICTP, Trieste preprint IC-92-129, July 1992;
H.O. Girotti and H.J. Rothe, Z. Phys. {\bf C27} (1985) 559;
K. Haller, Phys. Rev. {\bf D36} (1987) 1830, 1839.}

\refout
\endpage

\end